\documentstyle[sprocl]{article}

\def\be{\begin{equation}}
\def\ee{\end{equation}}
\def\bea{\begin{eqnarray}}
\def\eea{\end{eqnarray}}
\input epsf

\begin{document}

\title{Cranked relativistic Hartree-Bogoliubov theory:
\\ Superdeformation in the $A\sim 190$ mass region}

\author{A.\ V.\ Afanasjev$^{1,2}$
and P.\ Ring$^1$}

\address{$^1$Physik-Department der Technischen Universit{\"a}t
M{\"u}nchen,\\ D-85747 Garching, Germany \\
$^2$Laboratory of Radiation Physics, Institute of Solid State 
Physics,\\ University of Latvia, LV 2169 Salaspils, Miera str. 
31, Latvia\\E-mail: 
Anatoli.Afanasjev@Physik.TU-Muenchen.DE} 

\maketitle\abstracts{A systematic investigation of the 
yrast superdeformed (SD) rotational bands in even-even nuclei 
of the $A\sim 190$ mass region has been performed within the 
framework of the cranked relativistic Hartree-Bogoliubov 
theory. The particle-hole channel of this theory is treated
fully relativistically, while a finite range two-body force 
of Gogny type is used in the particle-particle (pairing)
channel. Using the well established parameter sets NL1 for
the Lagrangian and D1S for the Gogny force, very good 
description of experimental data is obtained with no
adjustable parameters.}


 Despite the fact that superdeformation at high spin has been 
studied experimentally and theoretically for one and half decade 
a number of theoretical questions 
such as, for example, the underlying mechanism of identical 
bands and the role of pairing correlations in the regime
of weak pairing in rotating nuclei, remains still not fully 
resolved and further development of theoretical tools is 
definitely required. During the last decade the relativistic
mean field (RMF) theory \cite{R.96} became a standard microscopic 
tool of nuclear structure studies. Systematic investigation 
of SD rotating nuclei in the regime of weak 
pairing correlations in the $A\sim 150$
\cite{KR.93,AKR.96,Hung,ALR.98} 
and $A\sim 60$ (see Ref.\ \cite{A60} and 
references therein) mass regions revealed that cranked 
relativistic mean field (CRMF) theory \cite{KR,AKR.96}, in 
which pairing correlations are neglected, provides an 
astonishingly accurate description of the properties of 
SD bands, such as moments of inertia, transition quadrupole 
moments $Q_t$, effective alignments $i_{eff}$, single-particle 
properties at superdeformation etc. In particular we have to
keep in mind that this theory has only seven free parameters fitted 
to the properties of few spherical nuclei \cite{NL1}.

\begin{figure}[t]
\epsfxsize 12.0cm
\epsfbox{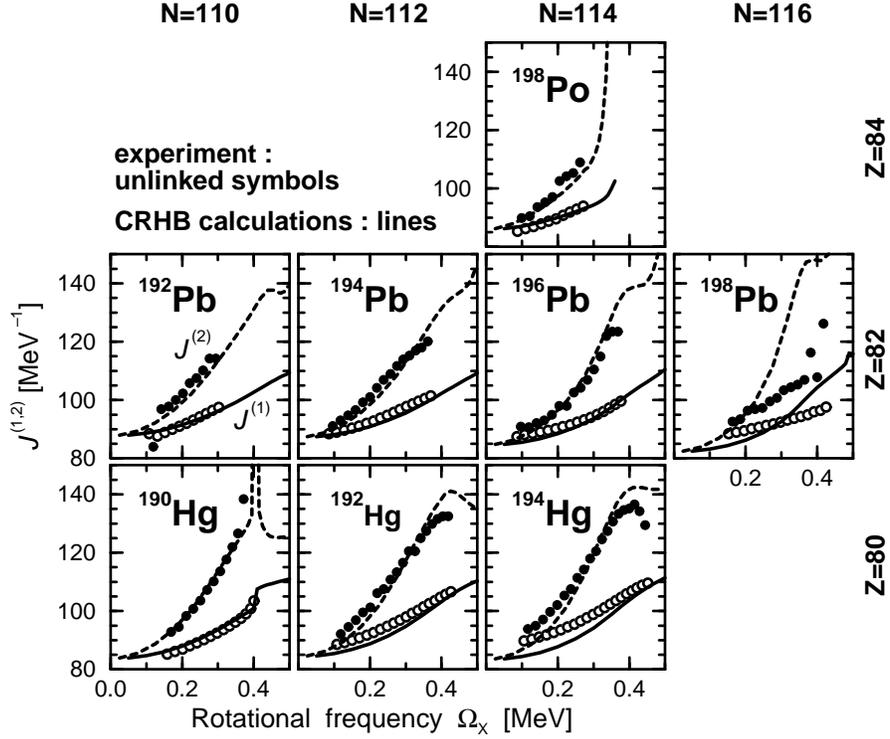}
\caption{ Experimental and calculated kinematic ($J^{(1)}$) and 
dynamic ($J^{(2)}$) moments of inertia of the yrast SD bands 
in even-even nuclei of the $A\sim 190$ mass region.
Experimental $J^{(1)}$ and $J^{(2)}$ moments of 
inertia are shown by open and solid circles, 
respectively. Solid and dashed lines are used for the 
$J^{(1)}$ and $J^{(2)}$ moments obtained in the CRHB 
calculations.}
\label{sysj2j1}
\end{figure}

 One should clearly recognize that the neglect of pairing
correlations used in CRMF theory is an approximation 
because pairing correlations being weak are still present
even at the highest rotational frequencies. Moreover, the
rotational properties of nuclei at low and medium spin
are strongly affected by pairing correlations. In order 
to describe such properties within the relativistic 
framework, cranked relativistic Hartree-Bogoliubov theory 
has been developed \cite{A190,CRHB,J1Rare}. This theory 
is an extension of CRMF theory to the description of pairing 
correlations in the rotating frame. In this theory the particle-hole
channel is treated fully relativistically on the Hartree level,
while the particle-particle channel is approximated by the
best currently available non-relativistic interaction: the
pairing part of the Gogny force. The use of  this force has 
a clear advantage since it provides both an automatic cutoff 
of high-momentum components and, as follows from non-relativistic 
studies, an excellent description of pairing properties in finite 
nuclei. An additional feature of CRHB theory is that approximate
particle number projection is performed by means of the 
Lipkin-Nogami method (further APNP(LN)) \cite{L.60,N.64,PNL.73}. 
The comparative study of pairing and rotational properties in the rare
earth region performed within the frameworks of CRHB theory
and non-relativistic cranked Hartree-Fock-Bogoliubov theory
based on the finite range force of the Gogny type \cite{J1Rare}
indicates that APNP(LN) plays a more important role in the 
relativistic calculations most likely reflecting the lower 
effective mass.

In Refs.\ \cite{A190,CRHB} CRHB theory has been applied for a 
systematic investigation of the properties of SD bands in 
even-even nuclei of the $A\sim 190$ mass region.
The calculations have been performed using the well 
established parameter sets NL1 \cite{NL1} for the RMF Lagrangian and 
D1S \cite{D1S} for the Gogny force. Fig.\ \ref{sysj2j1} compares 
the experimental dynamic and 
kinematic moments of inertia with the ones obtained in the
CRHB calculations. Since the SD bands in $^{190,192}$Hg, 
$^{196,198}$Pb and $^{198}$Po are not linked to the low
spin level scheme, their 'experimental' spin values and 
thus kinematic moments of inertia have been established
based on the comparison with calculated kinematic moments 
of inertia. Note that the analysis of experimental and
calculated effective alignments $i_{eff}$ between the bands 
in different nuclei confirms the present assignment
of the spin values for unlinked bands. One can
see that very good agreement exists in all the 
cases with an exception of the yrast SD band in 
$^{198}$Pb. The investigation of the structure
of the SD bands in neighboring odd nuclei is needed 
for a better understanding of the problems seen in this nucleus.

\begin{figure}[h]
\epsfxsize 12.0cm
\epsfbox{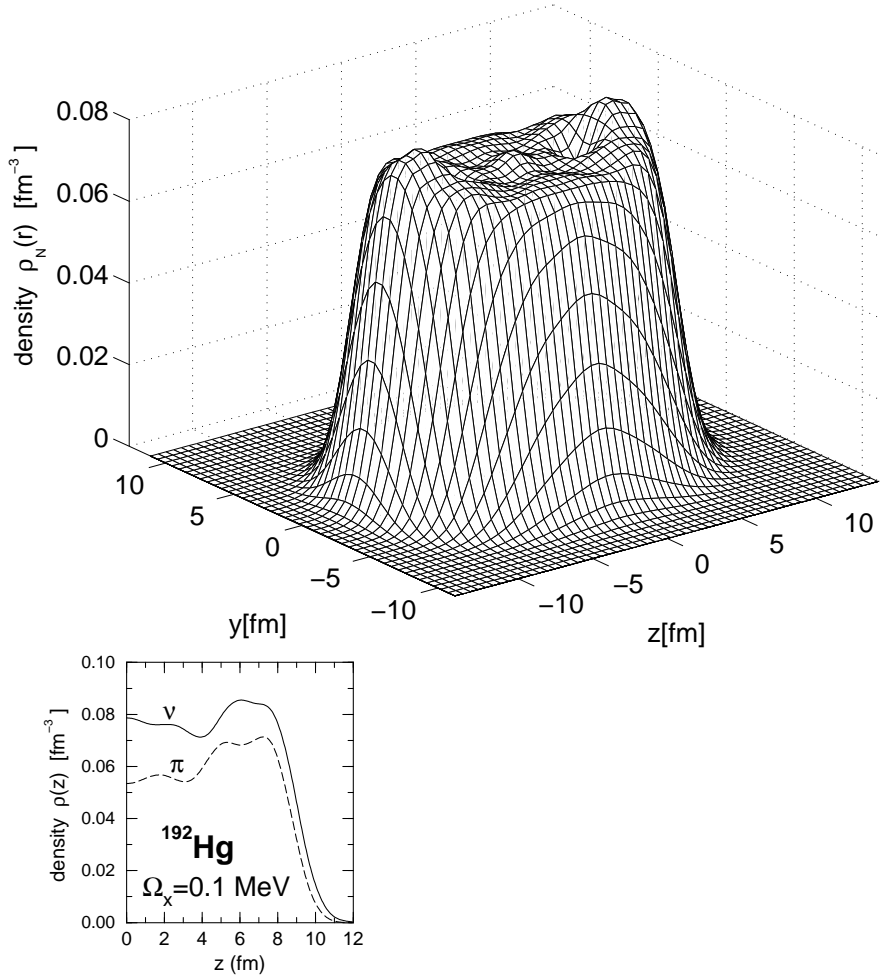}
\caption{The self-consistent scalar neutron density
as a function of the $y$- and $z$-coordinates at the value 
$x=0.308$ fm for the yrast SD band in $^{192}$Hg at rotational 
frequency $\Omega_x=0.1$ MeV (top panel).
The self-consistent scalar neutron and proton densities
as a function of the $z$-coordinate at the values $x=y=0.308$ fm
(bottom panel). Note that the symmetry axis points in $z$-direction, 
while the $x$-axis is the axis of rotation.}
\label{dens}
\end{figure}

Proton and neutron scalar density distributions for the
yrast SD band in $^{192}$Hg calculated at rotational frequency
$\Omega_x=0.1$ MeV are shown in Fig.\ \ref{dens}. Both distributions 
show considerable variations as a function of the coordinate. These 
variations are caused by shell effects. It is interesting to mention 
that the maximal density is reached along the symmetry axis at the 
distance of $6-8$ fm from the center of nucleus. These density
distributions correspond to a transition quadrupole moment
$Q_t=19.6$ $e$b. A detailed comparison presented in Ref.\ \cite{A190}
shows that the results of the CRHB calculations are within the
error bars of available experimental data on transition quadrupole
moments $Q_t$ for yrast SD bands of even-even nuclei in the $A\sim 190$ 
mass region.

 In conclusion, cranked relativistic Hartree-Bogoliubov
theory has been developed and applied for a systematic investigation of 
SD bands of even-even nuclei in the $A\sim 190$ mass region. 
Using well established parameter sets for the RMF Lagrangian and 
Gogny force the available experimental data is described very well 
without any new adjustable parameters. Further investigations of odd
and odd-odd nuclei are needed for a deeper understanding of
the properties of SD bands in CRHB theory. Such an investigation 
is in progress.

\section*{Acknowledgments}
  A.V.A. acknowledges support from the Alexander von Humboldt
Foundation. This work is also supported in part by the
Bundesministerium f{\"u}r Bildung und Forschung under the
project 06 TM 979.

\section*{References}
\begin{thebibliography}{99}

\bibitem{R.96} P. Ring, {\it Prog.\ Part.\ Nucl.\ Phys.} {\bf 37}, 193
(1996).

\bibitem{KR.93} J. K{\"o}nig and P. Ring, {\it Phys.\ Rev.\ Lett.}
{\bf 71}, 3079 (1993).
 
\bibitem{AKR.96} A.V. Afanasjev, J. K{\"o}nig and P. Ring,
{\it Nucl.\ Phys.} A {\bf 608}, 107 (1996).

\bibitem{Hung} A.V. Afanasjev, G.A. Lalazissis and P. Ring,
{\it Acta Phys.\ Hung.} {\bf 6}, 299 (1997).

\bibitem{ALR.98} A.V. Afanasjev, G.A. Lalazissis and P. Ring,
{\it Nucl.\ Phys.} {\bf 634}, 395 (1998).

\bibitem{A60} A.V. Afanasjev, I. Ragnarsson and P. Ring,
{\it Phys.\ Rev.} C {\bf 59}, 3166 (1999).

\bibitem{KR} W. Koepf and P. Ring, {\it Nucl.\ Phys.}
A {\bf 493}, 61 (1989); {\bf 511}, 279 (1990).


\bibitem{NL1} P.-G. Reinhard, M. Rufa, J. Maruhn,
W. Greiner and J. Friedrich, {\it Z.\ Phys.}  A {\bf 323},
13 (1986).

\bibitem{A190} A.V. Afanasjev, J. K\"onig and P. Ring,
{\it Phys.\ Rev.} {\bf C 60}, 051303 (1999).

\bibitem{CRHB} A.V. Afanasjev, P. Ring and
J. K\"onig, {\it Nucl.\ Phys.} A, in press (see
also report nucl-th/0001054).

\bibitem{J1Rare} A.V. Afanasjev, J. K\"onig, P. Ring, 
L.M. Robledo and J.L. Egido, {\it  Phys.\ Rev.} C, in press

\bibitem{L.60} H.J. Lipkin, {\it Ann.\ Phys.} {\bf 31}, 525 
(1960).

\bibitem{N.64} Y. Nogami, {\it Phys.\ Rev.} {\bf 134}, 313 (1964).

\bibitem{PNL.73} H.C. Pradhan, Y. Nogami, and J. Law, 
{\it Nucl.\ Phys.} {\bf A201}, 357 (1973).

\bibitem{D1S} J.F. Berger, M. Girod and D. Gogny,
{\it Comp.\ Phys.\ Comm.}  {\bf 63}, 365 (1991).

\end{thebibliography}
\end{document}